\documentclass{snmp2003}

\begin{document}

\FirstPageHeading{Kirilyuk}

\ShortArticleName{Universal Symmetry of Complexity} 

\ArticleName{Universal Symmetry of Complexity\\
 and Its Manifestations at Different Levels\\
 of World Dynamics}

\Author{Andrei P. KIRILYUK} \AuthorNameForHeading{A.P. Kirilyuk}
\AuthorNameForContents{KIRILYUK A.P.}
\ArticleNameForContents{Universal Symmetry of Complexity and Its
Manifestations at Different Levels of World Dynamics}
\Address{Institute of Metal Physics, pr. Vernadskogo 36, Kyiv
03142, Ukraine} \Email{kiril@ metfiz.freenet.kiev.ua}

\Abstract{The unreduced, universally nonperturbative analysis of
arbitrary interaction process, described by a quite general
equation, provides the truly complete, ``dynamically multivalued"
general solution that leads to dynamically derived, universal
definitions of randomness, probability, chaoticity, complexity,
fractality, self-organisation, and other properties, extending
their axiomatic introduction in the conventional, dynamically
single-valued theory. Any real system emergence, structure, and
behaviour can be expressed now by the universal law of
conservation, or symmetry, of complexity that unifies extended
versions of any (correct) symmetry, law, or ``principle".
Particular applications of the universal symmetry of complexity,
from fundamental physics to biology and theory of consciousness,
provide old mysteries solutions and new research perspectives.}

\section{Introduction: The unreduced symmetry of nature}

The \emph{conventional symmetry} idea used in physics and
mathematics (see e. g.
\cite{Kirilyuk:bluman&kumei,Kirilyuk:fushchich&shtelen&serov,Kirilyuk:elliott&dawber})
assumes the existence of \emph{previously determined} (eventually
postulated) structures and properties that appear as dynamical
laws relating ``variables" $\cal X$ and ``parameters" $\cal P$:
$\cal C(X;P) \rm =0$. A formal symmetry, or ``invariance" of the
law, is introduced usually through the symmetry transformation
``operator", $\cal \hat S$, whose action is compatible with the
law in question, $\cal \hat S[C(X;P)] \rm =0$. This fact can be
used to reduce the law expression to a desired particular form,
including the explicit, functional relation between
\emph{separated} variables and parameters, or ``(exact) solution"
of a problem: $\cal {X} \mit =f \cal (P)$. However, application of
this symmetry concept to the real world dynamics reveals
irreducible \emph{ruptures} between various known symmetries and
systematic \emph{violations} of almost every exact symmetry,
accounting for the real world irregularity, which leads to the
``(spontaneously) broken symmetry" concept with quite fundamental
consequences.

A much more general interpretation of symmetry is possible
\cite{Kirilyuk:Kirilyuk97,Kirilyuk:Kirilyuk02}, within which the
\emph{real} world structure \emph{explicitly emerges} as
\emph{inevitable realisation} of the \emph{universal dynamic
symmetry}, or ``conservation law", $C= \rm const$, where
\emph{all} observed entities, properties, and measured quantities
are \emph{derived} as forms or manifestations of that universal
symmetry, remaining thus \emph{always exact} (unbroken), but
producing all the observed \emph{irregularities}. \emph{Any}
(correct) dynamical law, $\cal C(X;P)=C_{\rm 0}$, as well as the
\emph{unreduced} problem solution, $\cal X =F(P)$, is also
obtained now, in its essentially extended, \emph{causally
complete} form, as a \emph{rigorously derived} (rather than
postulated), \emph{totally realistic} manifestation of that
single, \emph{intrinsically unified} symmetry. We show that the
role of such unified symmetry belongs uniquely to the
\emph{universal symmetry (conservation) of dynamic complexity},
where the latter quantity is rigorously derived from the
\emph{unreduced} (nonperturbative) analysis of arbitrary (real)
\emph{interaction process}
\cite{Kirilyuk:Kirilyuk97,Kirilyuk:Kirilyuk02}. This unreduced
complexity, giving rise to the unified dynamical ``order of the
universe", is essentially different from conventional complexity
versions which do not originate from the unreduced problem
solution, but reflect a strongly reduced (zero-dimensional)
\emph{projection} of real system dynamics.

\section{Unreduced dynamic complexity of arbitrary interaction\\
process, its conservation and internal transformation}

Mathematical expression of dynamics of a vast variety of
interaction processes can be generalised in the form of
``existence equation"
\cite{Kirilyuk:Kirilyuk97,Kirilyuk:Kirilyuk02,Kirilyuk:Kirilyuk96,Kirilyuk:Kirilyuk02a}
that actually only fixes the fact of interaction within a system
of a given configuration:
\begin{equation}\label{Kirilyuk:eq1}
\left[ {h_{\rm g} \left( \xi  \right) + V_{{\rm eg}} \left( {q,\xi
} \right) + h_{\rm e} \left( q \right)} \right]\mit\Psi \left(
{q,\xi } \right) = E\Psi \left( {q,\xi } \right),
\end{equation}
where $\mit\Psi(q,\xi)$ is the system ``state-function", which
totally determines its configuration and depends on the degrees of
freedom, $\xi$ and $q$, of the system components, $h_{\rm g}(\xi)$
and $h_{\rm e}(q)$ are ``generalised Hamiltonians" of the free
(non-interacting) components (i. e. measurable functions
eventually expressing dynamic complexity, defined below), $V_{{\rm
eg}}(q,\xi)$ is (arbitrary) interaction potential, $E$ is the
generalised Hamiltonian eigenvalue for the whole system, and any
number of interacting components can actually be implied behind
equation (\ref{Kirilyuk:eq1}), leading to the same results below
\cite{Kirilyuk:Kirilyuk02}.

It will be convenient to express the problem in terms of the
internal system states by performing expansion of the
state-function $\mit\Psi(q,\xi)$ over the complete system of
eigenfunctions, $\{\phi _n (q)\}$, for all free-state degrees of
freedom but one (described here by the $\xi$ variable and usually
representing the global system configuration, such as spatial
coordinates of its structure):
\begin{equation}\label{Kirilyuk:eq2}
\mit\Psi \left( {q,\xi } \right) = \sum\limits_n {\psi _n \left(
\xi \right)} \phi _n \left( q \right),\qquad h_{\rm e} \left( q
\right)\phi _n \left( q \right) = \varepsilon _n \phi _n \left( q
\right).
\end{equation}
Substituting expansion (\ref{Kirilyuk:eq2}) into the existence
equation (\ref{Kirilyuk:eq1}), multiplying it by $\phi _n^ * (q)$,
integrating over $q$ variables (or using other ``scalar product"
definition), and assuming the orthonormality of eigenfunctions
$\{\phi _n(q)\}$, we get a system of equations, which is
equivalent to the starting existence equation and includes all its
particular cases (e. g. nonlinear or time-dependent forms):
\begin{gather}
\left[ {h_{\rm g} \left( \xi  \right) + V_{00} \left( \xi \right)}
\right]\psi _0 \left( \xi  \right) + \sum\limits_n {V_{0n} }
\left( \xi  \right)\psi _n \left( \xi  \right) = \eta \psi _0
\left( \xi  \right),\nonumber\\
\left[ {h_{\rm g} \left( \xi \right) + V_{nn} \left( \xi \right)}
\right]\psi _n \left( \xi \right) + \sum\limits_{n' \ne n}
{V_{nn'} } \left( \xi \right)\psi _{n'} \left( \xi  \right) = \eta
_n \psi _n \left( \xi \right) - V_{n0} \left( \xi \right)\psi _0
\left( \xi  \right),\label{Kirilyuk:eq3}
\end{gather}
where $\eta _n  \equiv E - \varepsilon _n$,
\begin{equation*}
V_{nn'} \left( \xi  \right) = \int\limits_{\mit \Omega _q } {dq}
\phi _n^ *  \left( q \right)V_{{\rm eg}} \left( {q,\xi }
\right)\phi _{n'} \left( q \right),
\end{equation*}
and we have separated the equation with $n = 0$ from the system
(\ref{Kirilyuk:eq3}), so that other $n \ne 0$ (also below) and
$\eta \equiv \eta _0$.

Expressing $\psi_n(\xi)$ from equations (\ref{Kirilyuk:eq3})
through $\psi_0(\xi)$ by the standard Green function technique
\cite{Kirilyuk:Kirilyuk92,Kirilyuk:Dederichs} and inserting the
result into the equation for $\psi_0(\xi)$, we restate the problem
in terms of \emph{effective existence equation}, formally
involving only the selected degrees of freedom $\xi$:
\begin{equation}\label{Kirilyuk:eq4}
\left[ {h_{\rm g} \left( \xi  \right) + V_{{\rm eff}} \left( {\xi
;\eta } \right)} \right]\psi _0 \left( \xi  \right) = \eta \psi _0
\left( \xi  \right),
\end{equation}
where the \emph{effective (interaction) potential (EP)}, $V_{{\rm
eff}}(\xi;\eta)$, is given by

\begin{gather}
V_{{\rm eff}} \left( {\xi ;\eta } \right) = V_{00} \left( \xi
\right) + \hat V\left( {\xi ;\eta } \right),\qquad \hat V\left(
{\xi ;\eta } \right)\psi _0 \left( \xi  \right) =
\int\limits_{\mit \Omega _\xi  } {d\xi 'V\left( {\xi ,\xi ';\eta }
\right)} \psi _0 \left(
{\xi '} \right),\nonumber\\
V\left( {\xi ,\xi ';\eta } \right) = \sum\limits_{n,i}
{\frac{{V_{0n} \left( \xi  \right)\psi _{ni}^0 \left( \xi
\right)V_{n0} \left( {\xi '} \right)\psi _{ni}^{0*} \left( {\xi '}
\right)}}{{\eta  - \eta _{ni}^0  - \varepsilon _{n0} }}}\ ,\qquad
\varepsilon _{n0}  \equiv \varepsilon _n  - \varepsilon _0\ ,
\label{Kirilyuk:eq5}
\end{gather}
and $\{\psi_{ni}^0(\xi)\}$, $\{\eta_{ni}^0\}$ is the complete set
of eigenfunctions and eigenvalues for an auxiliary, truncated
system of equations (where $n,n' \ne 0$):
\begin{equation}\label{Kirilyuk:eq6}
\left[ {h_{\rm g} \left( \xi  \right) + V_{nn} \left( \xi \right)}
\right]\psi _n \left( \xi  \right) + \sum\limits_{n' \ne n}
{V_{nn'} \left( \xi  \right)} \psi _{n'} \left( \xi  \right) =
\eta _n \psi _n \left( \xi  \right).
\end{equation}
The general solution of the initial existence equation
(\ref{Kirilyuk:eq1}) is then obtained as
\cite{Kirilyuk:Kirilyuk97,Kirilyuk:Kirilyuk02,Kirilyuk:Kirilyuk96,
Kirilyuk:Kirilyuk02a,Kirilyuk:Kirilyuk92}:
\begin{gather}
{\mit \Psi} \left( {q,\xi } \right) = \sum\limits_i {c_i } \left[
{\phi _0 \left( q \right) + \sum\limits_n {\phi _n } \left( q
\right)\hat g_{ni} \left( \xi  \right)} \right]\psi _{0i} \left(
\xi  \right),\nonumber\\
\psi _{ni} \left( \xi  \right) = \hat g_{ni} \left( \xi
\right)\psi _{0i} \left( \xi  \right) \equiv \int\limits_{\mit
\Omega _\xi  } {d\xi 'g_{ni} \left( {\xi ,\xi '} \right)\psi _{0i}
\left({\xi '} \right)},\nonumber\\
g_{ni} \left( {\xi , \xi '} \right) = V_{n0} \left( {\xi '}
\right)\sum\limits_{i'} {\frac{{\psi _{ni'}^0 \left( \xi
\right)\psi _{ni'}^{0*} \left( {\xi '} \right)}}{{\eta _i  - \eta
_{ni'}^0  - \varepsilon _{n0} }}}\ ,\label{Kirilyuk:eq7}
\end{gather}
where $\{\psi_{0i}(\xi)\}$ are the eigenfunctions and $\{\eta_i\}$
the eigenvalues found from equation (\ref{Kirilyuk:eq4}), while
the coefficients $c_i$ should be determined by state-function
matching on the boundary where the effective interaction vanishes.
The observed system density, $\rho(q,\xi)$, is given by the
squared modulus of the state-function, $\rho(q,\xi) = \left|{\mit
\Psi}(q,\xi)\right|{}^2$ (for ``quantum" and other ``wave-like"
levels of complexity), or by the state-function itself,
$\rho(q,\xi) = {\mit \Psi}(q,\xi)$ (for ``particle-like" levels)
\cite{Kirilyuk:Kirilyuk97}.

Although the ``effective" problem formulation of equations
(\ref{Kirilyuk:eq4})-(\ref{Kirilyuk:eq7}) forms the basis of the
well-known optical, or effective, potential method (see e. g.
\cite{Kirilyuk:Dederichs}), it is actually used in its reduced,
perturbative versions, where the ``nonintegrable", nonlinear links
in the above EP and state-function expressions are cut in exchange
to the closed, ``exact" solution. However, this reduction kills
the \emph{essential, dynamic nonlinearity} of the real system,
together with its intrinsic complexity and chaoticity, and thus
replaces the natural symmetry of complexity by an artificial,
simplified symmetry of perturbative solutions
\cite{Kirilyuk:Kirilyuk97,Kirilyuk:Kirilyuk02,Kirilyuk:Kirilyuk96,
Kirilyuk:Kirilyuk02a,Kirilyuk:Kirilyuk92}. Indeed, it is not
difficult to show that the unreduced ``effective" problem has
\emph{many} locally complete and therefore \emph{incompatible}
solutions, \emph{each} of them being equivalent to the
\emph{single}, ``complete" solution of the reduced problem,
usually attributed also to the initial formulation of equations
(\ref{Kirilyuk:eq1}), (\ref{Kirilyuk:eq3}). If $N_\xi$ and $N_q$
are the numbers of terms in sums over $i$ and $n$ in equation
(\ref{Kirilyuk:eq5}), then the total number of eigenvalues of
equation (\ref{Kirilyuk:eq4}) is $N_{\max }  = N_\xi(N_\xi N_q +
1) = (N_\xi)^2 N_q  + N_\xi$, which gives the $N_\xi$-fold
redundance of the usual ``complete" set of $N_\xi N_q$
eigen-solutions of equations (\ref{Kirilyuk:eq3}) plus an
additional, ``incomplete" set of $N_\xi$ solutions. Each redundant
solution, \emph{intrinsically unstable} with respect to system
transitions to other solutions, can be called system
\emph{realisation}, since it represents a completely determined
system configuration. The total number of ``regular", complete
system realisations is $N_\Re = N_\xi$, whereas the mentioned
additional set of solutions forms a special, ``intermediate"
realisation that plays the role of transitional state during
system jumps between the regular realisations and provides thus
the universal, \emph{causally complete} extension of the quantum
\emph{wavefunction} and classical \emph{(probability) distribution
function} \cite{Kirilyuk:Kirilyuk97,Kirilyuk:Kirilyuk02}.

Thus rigorously derived, qualitatively new property of
\emph{dynamic multivaluedness} of the unreduced problem solution
is confirmed by its ``geometric" analysis and particular
applications
\cite{Kirilyuk:Kirilyuk97,Kirilyuk:Kirilyuk02,Kirilyuk:Kirilyuk96,
Kirilyuk:Kirilyuk02a,Kirilyuk:Kirilyuk92}. It provides the
intrinsic, omnipresent, and irreducible source of \emph{purely
dynamic, or causal, randomness}: the incompatible system
realisations, being \emph{equally} real, should \emph{permanently}
replace one another, in a \emph{causally random} order, so that
the observed density of any real system should be presented as the
\emph{dynamically probabilistic} sum of the individual realisation
densities, $\{\rho _r(\xi,q)\}$, obtained by solution of the
effective existence equation (\ref{Kirilyuk:eq4}):
\begin{equation}\label{Kirilyuk:eq8}
\rho \left( {\xi ,Q} \right) = \sum\limits_{r = 1}^{N_\Re  } {^
\oplus  \rho _r \left( {\xi ,Q} \right)} ,
\end{equation}
where summation is performed over all (observable) system
realisations, numbered by $r$, and the sign $\oplus$ serves to
designate the special, dynamically probabilistic meaning of the
sum derived above and consisting in \emph{permanent change} of
regular realisations in \emph{dynamically random (chaotic)} order
by transition through the intermediate realisation. The
dynamically obtained, \emph{a priori probability} of the $r$-th
realisation emergence, $\alpha _r$, is determined, in general, by
the number, $N_r$, of elementary, experimentally unresolved
realisations it contains:
\begin{equation}\label{Kirilyuk:eq9}
\alpha _r \left( {N_r } \right) = \frac{{N_r }}{{N_\Re }}\quad
\left( {N_r  = 1, \ldots ,N_\Re  ;\ \sum\limits_r {N_r }  = N_\Re
} \right),\qquad \sum\limits_r {\alpha _r }  = 1\ .
\end{equation}
According to the ``generalised Born's rule", obtained by dynamical
matching in the intermediate realisation (wavefunction) phase, the
dynamic probability values are determined by the generalised
wavefunction obeying the causally derived, \emph{universal
Schr\"odinger equation}
\cite{Kirilyuk:Kirilyuk97,Kirilyuk:Kirilyuk02} (see below).

Another important property of the unreduced solution, closely
related to the above dynamic multivaluedness, is \emph{dynamic
entanglement} between the interacting entities (degrees of
freedom) within each realisation, which appears as dynamically
weighted products of functions of $\xi$ and $q$ in equations
(\ref{Kirilyuk:eq7}) and determines the \emph{tangible new
quality} of the emerging interaction results. It leads to the
\emph{dynamical} system squeeze, or \emph{reduction}, or
\emph{collapse}, to the emerging configuration of each
realisation, alternating with the reverse \emph{dynamic
disentanglement}, or \emph{extension}, of interacting entities to
a quasi-free state in the intermediate realisation (wavefunction),
during transitions between realisations
\cite{Kirilyuk:Kirilyuk97,Kirilyuk:Kirilyuk02,Kirilyuk:Kirilyuk96,
Kirilyuk:Kirilyuk02a}. The dynamically multivalued entanglement is
a totally autonomous \emph{process}, driven only by the system
interaction and characterised by the intrinsic
\emph{nonseparability} and \emph{irreversible} direction.
Nonseparable component entanglement gives rise to the explicitly
emerging, physically real \emph{space} (in the form of the
squeezed, final realisation configuration, or generalised space
``point"), while the irreversible, unceasing and \emph{spatially
chaotic} realisation change determines the causal \emph{time} flow
\cite{Kirilyuk:Kirilyuk97,Kirilyuk:Kirilyuk02}. These properties
of the dynamically multivalued entanglement between the
interacting components are hierarchically reproduced and amplified
within the \emph{dynamically fractal} structure of the unreduced
problem solution, which can be obtained by application of the same
EP method to the truncated system of equations
(\ref{Kirilyuk:eq6}) whose solutions are used in the expressions
of the first level of solution, equations (\ref{Kirilyuk:eq5}),
(\ref{Kirilyuk:eq7}). We obtain thus the causally complete
extension of the conventional, dynamically single-valued
fractality and the true meaning of (\emph{any} real) system
\emph{nonintegrability}, which takes the form of the permanently
changing, dynamically probabilistic (``living") fractal hierarchy
of the unreduced problem solution, possessing the rigorously
obtained properties of \emph{explicit structure emergence
(creativity)} and \emph{dynamic adaptability} (self-consistent
configuration of the ``effective" solution of equations
(\ref{Kirilyuk:eq4})-(\ref{Kirilyuk:eq7}))
\cite{Kirilyuk:Kirilyuk97,Kirilyuk:Kirilyuk02,
Kirilyuk:Kirilyuk02a}.

Now that the \emph{dynamically multivalued} structure of the
unreduced interaction process has been \emph{explicitly revealed},
we can provide the unrestricted, \emph{universally applicable}
definition of \emph{dynamic complexity}, $C$, as any growing
function of realisation number, $C = C(N_\Re), \ {{dC}
\mathord{\left/ {\vphantom {{dC} {dN_\Re   > 0}}} \right.
 \kern-\nulldelimiterspace} {dN_\Re   > 0}}$, or the rate
of their change, equal to zero for the (unrealistic) case of only
one realisation, $C(1) = 0$. It is just the latter,
unrealistically simplified ``model" (zero-dimensional, point-like
projection) of reality which is \emph{exclusively} considered in
the conventional, dynamically single-valued, or \emph{unitary},
theory, including its concepts of ``complexity", ``chaoticity",
``self-organisation", etc., which explains all its persisting
``mysteries" and ``difficult" problems, easily finding their
dynamically multivalued, causally complete solution within the
unreduced complexity concept
\cite{Kirilyuk:Kirilyuk97,Kirilyuk:Kirilyuk02,Kirilyuk:Kirilyuk96,
Kirilyuk:Kirilyuk02a,Kirilyuk:Kirilyuk92} that emerges thus as the
direct, qualitative \emph{extension} of the unitary knowledge
model to the dynamically multivalued reality. In particular, the
properties of dynamic multivaluedness and entanglement show that
\emph{chaoticity} is \emph{synonymous to complexity}, in their
unreduced, \emph{omnipresent} versions.

In that way the \emph{regular} and \emph{separated} symmetries of
the unitary model, \emph{always} (fortunately!) \emph{violated} in
the real world (cf. the concept of ``spontaneously" broken
symmetry), are replaced, in the unreduced description, by the
single, \emph{intrinsically unified, but diverse in
manifestations, irregular, but exact (never broken) symmetry of
complexity} \cite{Kirilyuk:Kirilyuk97,Kirilyuk:Kirilyuk02}.
Moreover, contrary to the artificially imposed, external origin of
the conventional symmetries, mechanistically added to the
postulated structures, properties, and ``principles", the
universal symmetry of complexity emerges as the unique source of
existence giving rise, through the explicitly obtained relations,
to \emph{all} real entities and (correct) laws, in their causally
extended, complex-dynamical (multivalued) version. At a given
level of system complexity (described by the above solution
(\ref{Kirilyuk:eq4})-(\ref{Kirilyuk:eq9})), this irreducibly
\emph{dynamic} symmetry appears as equivalence between all
(elementary) system realisations meaning their ``equal chances" to
emerge and permanent actual change, as reflected in the
probability expression (\ref{Kirilyuk:eq9}) (also in its relation
to the wavefunction values) and dynamically probabilistic sum of
the general solution, equation (\ref{Kirilyuk:eq8}). All
realisations \emph{differ} in their detailed structure and are
taken by the system in a truly random order (equations
(\ref{Kirilyuk:eq4})-(\ref{Kirilyuk:eq9})), thus reproducing the
\emph{real world irregularity}, but the resulting internally
\emph{irregular} symmetry between them is \emph{exact} as such
(unbroken) and can be expressed simply as fixed realisation number
for any given system (interaction process).

However, the universal symmetry of complexity does not stop there:
it involves a \emph{qualitative} change of the \emph{form} of
complexity that \emph{preserves} its total \emph{quantity}.
Namely, the potential, or ``hidden" (latent) form of complexity,
called \emph{dynamic information} (and generalising ``potential
energy"), is transformed into the explicit, ``unfolded" form of
\emph{dynamic entropy} (extending entropy concept to any process),
so that their sum, the total system complexity remains unchanged,
which gives rise to all emerging entities, their properties and
behaviour (reflected in particular ``laws" and ``principles")
\cite{Kirilyuk:Kirilyuk97,Kirilyuk:Kirilyuk02}. The basic origin
of that complexity transformation is revealed by the same,
unreduced interaction description, containing the \emph{explicit
emergence} of \emph{always internally chaotic} entities and their
interactions (given by higher, fine levels of the fractal
hierarchy of unreduced interaction development).

The length element, $\Delta x$, of a complexity level is obtained
from solution of the unreduced ``effective" equation
(\ref{Kirilyuk:eq4})-(\ref{Kirilyuk:eq5}) as the distance between
the centres of the neighbouring realisation eigenvalues, $\Delta x
= \Delta \eta _i^r$, while the time flow rate emerges as intensity
(specified as \emph{frequency}, $\nu$) of \emph{realisation
change}. Since the emerging space and time represent the two
basic, universal forms of complexity, its universal, natural
measure should be independently proportional to measures of space
and time. It is easy to see that such complexity measure is
provided by \emph{action} quantity acquiring thus its extended,
\emph{essentially nonlinear}, meaning: $\Delta \cal A \mit =  -
E\Delta t + p\Delta x$, where $\Delta x$, and $\Delta t = {1
\mathord{\left/ {\vphantom {1 \nu }} \right.
\kern-\nulldelimiterspace} \nu }$ are the above dynamic space and
time increments, $\Delta \cal A$ is the corresponding
complexity-action increment, while the coefficients, $E$ and $p$,
are identified as energy and momentum. The action value always
decreases ($\Delta \cal A \rm < 0$) and represents the dynamic
information, whereas complexity-entropy change is the quantity
opposite in sign, $\Delta S =  - \Delta \cal A \rm > 0$, leaving
their sum, the total complexity, unchanged, $C = \cal A \mit + S =
{\rm const}$. Dividing the differential expression of conservation
(symmetry) of complexity by $\Delta t\left| {_{x = {\rm const}} }
\right.$, we get the generalised Hamilton-Jacobi equation
\cite{Kirilyuk:Kirilyuk97,Kirilyuk:Kirilyuk02}:
\begin{equation}\label{Kirilyuk:eq10}
\frac{{\Delta \cal A}}{{\Delta t}}\left| {_{x = {\rm const}} }
\right. + H\left( {x,\frac{{\Delta \cal A}}{{\Delta x}}\left| {_{t
= {\rm const}} ,t} \right.} \right) = 0,
\end{equation}
where the \emph{Hamiltonian}, $H = H(x,p,t)$, considered as a
function of emerging space-structure coordinate $x$, momentum $p =
\left( {{{\Delta \cal A} \mathord{\left/ {\vphantom {{\Delta A}
{\Delta x}}} \right. \kern-\nulldelimiterspace} {\Delta x}}}
\right)\left| {_{t = {\rm const}} } \right.$, and time $t$,
expresses the implemented, entropy-like form of differential
complexity, $H = \left( {{{\Delta S} \mathord{\left/ {\vphantom
{{\Delta S} {\Delta t}}} \right. \kern-\nulldelimiterspace}
{\Delta t}}} \right)\left| {_{x = {\rm const}} } \right.$. Because
of a dynamically random order of emerging system realisations, the
total time derivative of action, or \emph{Lagrangian}, $L =
{{\Delta \cal A} \mathord{\left/ {\vphantom {{\Delta A} {\Delta t
= pv - H}}} \right. \kern-\nulldelimiterspace} {\Delta t = pv -
H}}$, should be negative (where $v = {{\Delta x} \mathord{\left/
{\vphantom {{\Delta x} {\Delta t}}} \right.
\kern-\nulldelimiterspace} {\Delta t}}$ is the global-motion
velocity), which provides the rigorously derived, \emph{dynamic}
expression of the ``arrow of time" orientation to growing entropy:
\begin{equation*}
L < 0 \quad \Rightarrow \quad E,H\left( {x,\frac{{\Delta \cal
A}}{{\Delta x}}\left| {_{t = {\rm const}} ,t} \right.} \right)
> pv \geq 0\ .
\end{equation*}

Realisation change process can be considered also as two adjacent
complexity sublevels whose conserved total complexity, $C$, equals
to the product of complexity-entropy of localised (regular)
realisations and ``potential" wavefunction complexity, $C = S\mit
\Psi = {\rm const}$, meaning that $\cal A \mit \Psi  =  - S\Psi =
{\rm const}$, where $\mit \Psi$ is the wavefunction. The total
complexity change between two transitional states equals to zero,
$\Delta({\cal A} {\mit \Psi}) = 0$, which expresses the physically
evident permanence of the \emph{unique} state of wavefunction, and
gives the generalised \emph{causal quantization rule}:
\begin{equation}\label{Kirilyuk:eq11}
\Delta {\cal A} =  - {\cal A}_0 \frac{{\Delta {\mit \Psi} }}{\mit
\Psi}\ ,
\end{equation}
where ${\cal A}_0$ is a characteristic action value (${\cal A}_0$
may contain also a numerical constant reflecting specific features
of a given complexity level). Using equation (\ref{Kirilyuk:eq11})
in equation (\ref{Kirilyuk:eq10}) we obtain the \emph{generalised
Schr\"odinger equation} for $\mit \Psi$ in the form
\cite{Kirilyuk:Kirilyuk97,Kirilyuk:Kirilyuk02}:
\begin{equation}\label{Kirilyuk:eq12}
{\cal A}_0 \frac{{\partial {\mit \Psi} }}{{\partial t}} = \hat
H\left( {x,\frac{\partial }{{\partial x}},t} \right){\mit \Psi}\ ,
\end{equation}
where the Hamiltonian operator, $\hat H$, is obtained from the
Hamiltonian function $H = H(x,p,t)$ of equation
(\ref{Kirilyuk:eq10}) with the help of the causal quantization
relation of equation (\ref{Kirilyuk:eq11}).

The generalised Hamilton-Schr\"odinger formalism, equations
(\ref{Kirilyuk:eq10})-(\ref{Kirilyuk:eq12}), is a universal
expression of the symmetry of complexity. Expanding the
Hamiltonian in equation (\ref{Kirilyuk:eq10}) in a power series of
momentum and action, one obtains a form of the universal
Hamilton-Schr\"odinger formalism that can be reduced to any usual,
``model" equation by series truncation
\cite{Kirilyuk:Kirilyuk97,Kirilyuk:Kirilyuk02},
\begin{equation*}
\frac{{\partial {\mit \Psi} }}{{\partial t}} +
\sum\limits_{\scriptstyle m = 0 \hfill \atop \scriptstyle n = 1
\hfill}^\infty  {h_{mn} \left( {x,t} \right)} \left[ {{\mit \Psi}
\left( {x,t} \right)} \right]^m \frac{{\partial ^n {\mit \Psi}
}}{{\partial x^n }} + \sum\limits_{m = 0}^\infty  {h_{m0} } \left(
{x,t} \right)\left[ {{\mit \Psi} \left( {x,t} \right)} \right]^{m
+ 1}  = 0\ ,
\end{equation*}
(here the expansion coefficients, $h_{mn}(x,t)$, can be arbitrary
functions), which confirms its universality and shows the genuine,
unified origin of model equations, semi-empirically guessed and
postulated in the unitary theory. All fundamental laws and
``principles" of the conventional science, such as relativity
(special and general), principle of entropy increase, principle of
least action, other ``variational" principles, can now be
obtained, in their \emph{causally extended, complex-dynamical}
versions, from the same unified law of conservation, or symmetry,
of complexity \cite{Kirilyuk:Kirilyuk97,Kirilyuk:Kirilyuk02}.
Note, in particular, that the universal complexity conservation,
realised by its \emph{unceasing transformation} from decreasing
dynamic information (action) to increasing entropy, provides a
remarkable unification of the universal, extended versions of
least action principle (conventional mechanics) and entropy
increase principle (``second law" of thermodynamics), which
reveals the true meaning and origin of those ``well-known" laws.
In a similar way, the ``quantum", ``classical", and
``relativistic" effects and types of behaviour are \emph{causally}
explained now as inevitable, and thus universally extendible,
manifestations of the unified symmetry of unreduced complexity
\cite{Kirilyuk:Kirilyuk97,Kirilyuk:Kirilyuk02,Kirilyuk:Kirilyuk00}.
The underlying complex (multivalued) dynamics specifies the
essential difference of the symmetry of complexity from its
conventional imitations: the formal ``operators" of the latter are
replaced in the former by actual realisation change and complexity
unfolding, just forming the real, creative world dynamics (cf.
section 1).

One should emphasize the importance of genuine, \emph{dynamically
emerging}, or ``essential" \emph{nonlinearity}, defined above and
closely related to the dynamic multivaluedness, for the universal
symmetry of complexity, as well as its fundamental difference from
the conventional, mechanistically defined (non-dynamic)
``nonlinearity". The essential nonlinearity \emph{inevitably}
emerges as a result of \emph{unreduced} interaction development,
even starting from a formally ``linear" initial problem
formulation (see equations
(\ref{Kirilyuk:eq1})-(\ref{Kirilyuk:eq7}) and the following
analysis). On the other hand, formally ``nonlinear" equations of
the standard approach, being analysed within its reduced, unitary
projection, cannot produce any truly new structure that would not
be actually postulated within the starting formulation, and
therefore they remain always \emph{basically linear}, as it is
confirmed by their invariably perturbative, or \emph{exact},
solutions. The \emph{real} nonlinearity appears as a dynamically
fractal network of self-developing interaction feedback loops,
explicitly revealed just in the ``effective" problem expression of
the generalised EP method
\cite{Kirilyuk:Kirilyuk97,Kirilyuk:Kirilyuk02,Kirilyuk:Kirilyuk96,
Kirilyuk:Kirilyuk02a,Kirilyuk:Kirilyuk92}. This \emph{emerging}
nonlinear structure forces the system to take, or ``collapse" to,
\emph{one} of its \emph{multiple} possible realisations, which
means that those realisations, \emph{actually} and
\emph{unceasingly} replacing one another in a \emph{dynamically
random}, or ``chaotic", order, are \emph{dynamically symmetric}
among them, while they always \emph{differ} in their detailed,
partially \emph{irregular} structure. However, the same system in
the phase of transition between its normal, ``localised"
realisations is forced, by the \emph{same} driving interaction, to
transiently disentangle its components up to their quasi-free
state of ``generalised wavefunction" (see above), and that's why
the system in this state temporarily behaves as a weekly
interacting, quasi-linear one. This remarkable, ``intermittent"
structure of unreduced interaction process, remaining totally
hidden in the dynamically single-valued projection of the
conventional theory, explains why and how the real system dynamics
naturally unifies the opposed, complementary properties of
quasi-linear and highly nonlinear behaviour and symmetry. The
realistic, causal explanation of ``wave-particle duality" and
``complementarity" in quantum systems (and classical
``distributed" systems as well) is only one particular consequence
of that omnipresent structure of the unreduced symmetry of
complexity
\cite{Kirilyuk:Kirilyuk97,Kirilyuk:Kirilyuk02,Kirilyuk:Kirilyuk96,
Kirilyuk:Kirilyuk02a,Kirilyuk:Kirilyuk92}.

It would be worthwhile to note finally that the described
conceptual transition from the conventional, dynamically
single-valued (unitary) to the proposed dynamically multivalued
(unreduced) description of system dynamics and the related upgrade
of the separated and broken unitary symmetries to the
intrinsically unified and exact symmetry of complexity involves
important progress in mathematical description of reality,
standing as the main, universal tool of science. We have seen that
the proposed advance in that description, which practically
totally eliminates the existing gap between real phenomena and
their unitary ``models", is realised simply due to the unreduced,
or \emph{really exact} mathematical analysis using quite ordinary
particular tools. This is certainly good news for mathematics,
which can thus preserve and develop its status as a universal
method and basis of \emph{objective} knowledge about reality, the
image that has considerably faded in the last period of growing
``uncertainty" \cite{Kirilyuk:Kline}, separations, and untractable
technical sophistication. On the other hand, the price that is
clearly to be paid for that essential and intrinsically
sustainable progress consists in the corresponding considerable,
fundamentally rooted upgrade of the scholar framework, which tends
traditionally to hide its real difficulties behind the externally
``solid" fa\c cade of the formally fixed ``existence and
uniqueness" theorems and other \emph{postulated} constructions.
This report presents a brief account of the means and results of
elementary realisation of that qualitative transition
demonstrating, in our opinion, both its feasibility and
inevitability in the future progress of science.

\section{Particular manifestations of the unified symmetry\\
of complexity}

We can only briefly outline here other manifestations of the
universal symmetry of complexity obtained for particular or
arbitrary levels of complexity and systems. One of them is
universal classification of all possible types of real system
behaviour which can vary continuously between the limiting cases
of \emph{uniform, or global, chaos} (quasi-homogeneous
distribution of probability for sufficiently different
realisations) and \emph{multivalued self-organisation, or
self-organised criticality} (inhomogeneous realisation probability
distribution, close elementary realisations)
\cite{Kirilyuk:Kirilyuk97,Kirilyuk:Kirilyuk02}. It is this latter
case that can be more successfully approximated by conventional,
dynamically single-valued (intrinsically regular) models, though
with irreducible fundamental losses (such as absence of
irreversible time flow). The universal criterion of transition
from self-organised (generally ordered) dynamics to the global
chaos, in both quantum and classical systems, is obtained in the
form of frequency resonance between interacting modes (such as
intra- and inter-component dynamics), which extends considerably
the concepts of both chaoticity and resonance
\cite{Kirilyuk:Kirilyuk97,Kirilyuk:Kirilyuk02,Kirilyuk:Kirilyuk96,
Kirilyuk:Kirilyuk02a,Kirilyuk:Kirilyuk92}. The observed
alternation of globally chaotic and self-organised levels in the
hierarchy of complexity is another manifestation of the universal
symmetry of complexity.

Application of the unreduced existence equation solution to the
simplest system of two attracting, initially homogeneous
protofields gives explicitly emerging field-particles, in the form
of spatially chaotic quantum beat processes, endowed with the
rigorously derived, realistic and unified versions of all
``mysterious" quantum features, ``relativistic" effects and
intrinsic properties (mass, electric charge, spin), obtained as
standard, inevitable manifestations of unreduced complexity
\cite{Kirilyuk:Kirilyuk97,Kirilyuk:Kirilyuk02,Kirilyuk:Kirilyuk00}.
The number (four), dynamic origin, properties and intrinsic
unification of fundamental interaction forces between particles
are obtained within the same picture. The true quantum chaos,
passing to classical chaos by the usual semiclassical transition,
intrinsically indeterminate quantum measurement, and dynamic
emergence of classical, permanently localised behaviour within a
closed, bound system (like atom) are obtained as naturally
emerging complexity levels, with important practical conclusions
for such popular applications as quantum computers,
nanotechnology, and quantum many-body systems with ``strong"
interaction
\cite{Kirilyuk:Kirilyuk97,Kirilyuk:Kirilyuk02,Kirilyuk:Kirilyuk96,
Kirilyuk:Kirilyuk00}. The obtained ``emergent" and causal world
picture includes also natural solution of the problems of
cosmology. Finally, symmetry of complexity manifestations for
biological and intelligent systems reveal the causal essence of
life, intelligence, and consciousness as high enough levels of
unreduced complexity, which leads to practically important
conclusions \cite{Kirilyuk:Kirilyuk97,Kirilyuk:Kirilyuk02,
Kirilyuk:Kirilyuk02a} and proves once more the universal
applicability of the unreduced symmetry of complexity.

\LastPageEnding

\end{document}